# A SIMPLE LAW FOR THE AVERAGE TIME HISTORY OF GAMMA-RAY BURSTS AND THEIR TIME DILATIONS

Boris E. Stern[1,2]

## ABSTRACT

Individual gamma ray bursts (GRBs) have very diverse time behavior - from a single pulse to a long complex sequence of chaotic pulses of different timescales. I studied light curves of GRBs using data from the *Compton Gamma Ray Observatory's* Burst and Transient Source Experiment (BATSE) and found that the average post-peak time history for a sample of 460 bursts obeys an unique and simple analytical law: $I \sim \exp(-(t/t_0)^{1/3})$ where $t$ is time measured from the peak of the event and $t_0$ is a constant ranging from 0.3 sec for strong bursts to $\sim 1$ sec for weak bursts. The average peak aligned profile follows this law with good accuracy in the whole range availible for analysis (from fractions of a second to $\sim 150$ seconds after the peak). Such a law with a single time constant characterising the overall sample of GRBs should have important physical meaning. The same analysis for solar flares gave a dependence close to $\exp(-(t/t_0)^{1/2})$ and some indication of a tendency to change the power index to 1/3 at $t > 50$ sec after the peak of the flare. The dependence of $t_0$ versus brightness of GRBs is presented. The fact that $t_0$ depends on the brightness apparantly confirms the recently discovered effect of time dilation of weak bursts which has a possible cosmological interpretation. The time dilation is detected at a confidence level of $7\sigma$ and it is slightly larger than was previously reported. The statistics (460 GRBs), however, still does not allow us to rule out an intrinsic anticorrelation between intensity and duration of GRBs as a possible explanation of the effect. Probably it is possible to confirm (or disprove) the cosmological origin of GRBs using the total existing statistics of strong bursts.

*Subject headings:* cosmology: theory – gamma rays: bursts

[1] Institute for Nuclear Research, Russian Academy of Sciences, Moscow, 117312, Russia, I: stern@inr.msk.su
[2] Stockholm Observatory, S-133 36 Saltsjöbaden, Sweden, I: stern@astro.su.se

## 1. Introduction

The characteristic distance to sources of gamma ray bursts is still unknown. A possibility that GBRs could originate from cosmological distances stimulated studies of their light curves in order to find signatures of the time dilation of weak GRBs relative to strong GRBs associated with a difference in their respective redshifts (Paczyński 1992; Piran 1992). These studies used different techniques. Davis et al. (1994) measured widths of individual pulses in multipeaked GRBs. They found a time dilation of order 2 for pulses in weak GRBs. Mitrofanov et al. (1994) compared average profiles of weak and strong GRBs and did not find any significant time dilation in a sample of 260 GRBs from the first BATSE catalog. Norris et al. (1994) using the same method as Mitrofanov et al. (1994) as well as other tests found a time dilation by a factor 2 using the larger statistics of the second BATSE catalog. Finally Norris et al. (1995) confirmed the time dilation effect measuring duration distributions of bright and dim GRBs.

All these studies did not consider possible characteristic shapes of measured distributions. One can make better estimates of the time dilation if one fits the data with a meaningful parametrized expression for the shape of distributions. In order to study possible signatures of cosmological effects in the time behaviour of GRBs, I decided to investigate the average shape of their light curves. To define the average time history one needs a time reference point for each GRB, the most reasonable being the time of the peak of the event. With this reference time one obtains an *average peak aligned profile* of GRBs, (Mitrofanov et al. 1994; Norris et al. 1994). I adopted the same approach and concentrated on the post-peak slope of the average profile as the pre-peak slope is influenced by the triggering effects and the corresponding data have a worse quality (the pre-peak slope is steeper, see Nemiroff et al. 1994). The purpose was to find an analytical expression that fitted the profile.

## 2. The average peak aligned profile of GRBs

All data utilised in this work were extracted from the publically available BATSE database in Goddard Space Flight Center. Time profiles of GBRs are derived from 64 ms resolution Large Area Detectors data by summing up counts in 4 discriminator channels covering the 25 keV - 1000 keV energy band. For relatively bright GRBs (peak count rates $I^p > 4000 \text{s}^{-1}$), positions of the peak were determined by the brightest 64 ms bin. For weaker events the peak was determined to be at the center of the 128, 256 or 512 ms brightest interval (in order of decreasing $I^p$).

To subtract the background I used linear fits – this procedure is not only simple but is also optimal because it preserves the statistics of weak tails of GBRs better than quadratic or higher order polynomial fits. To prepare the fitting procedure I made a visual examination of all GBR's time profiles determining fitting windows in a way to avoid regions with probable GBR contributions. For many events the 1024 ms resolution data covering a 470 s time interval were also used to determine two fitting windows – one before and one after a burst. Typically, the right fitting window was separated by at least 150 s from the burst peak (except for apparently short bursts).

After discarding all events with severe data gaps and a couple of events with a patalogically variable background I found 460 useful GBRs in the second BATSE catalog.

To divide GBRs into brightness groups I used 64 ms and 1024 ms integration time criteria for the peak brightness. This work is mainly based on the traditional 1024 ms brightness selection timescale because of its compromise between peak brightness and total energy release criteria. I did not select bursts using their durations as a criteria. The resulting profiles therefore include all GRBs, even those which are confined to within a single 64 ms bin. This is important because removing short GRBs from the sample would deform the average profile near its peak. Unresolved events give the proper normalization for the zero-point of the profile.

The average peak aligned time profiles obtained with this method extend for more than 150 seconds after the peak. I found that the shape of the average profile obeys with good precision a simple law:

$$I(t)/I^p = \exp(-(t/t_0)^{1/3})$$

where $t$ is the time after the peak, $I^p$ is the peak count rate (intensity) and $t_0$ is a constant. This is demonstrated in Figure 1 where the average time profiles for two brightness groups are plotted as $\log(I)$ vs $t^{1/3}$. The time constant $t_0$ depends on the brightness group ($t_0 = 0.48$ s for the bright group, $t_0 = 1.01$ s for the weak group) apparently demonstrating time dilation of weak bursts in approximate agreement with results of Norris et al. (1994, 1995).

The time profile averaged over 409 GBRs (the weakest events were excluded from this sample) has an intermediate $t_0$ ($t_0 = 0.72$ s) and closely follows the $\exp(-(t/t_0)^{1/3})$ law. However, if we believe in the cosmological nature of the time dilation it makes sense to rescale all light curves to the same redshift. This procedure allows us to get rid of a possible slight curvature resulting from mixing linear profiles of different slopes. I calculated the time profile averaged over 409 GRBs, where time profiles of 3 brightness groups covering the whole sample were rescaled to $t_0 = 1$. The residuals of the linear fit to this time profile are presented in Fig. 1b, while the time profile itself is presented in Fig. 2. The linear behavior is clearly visible up to $t \sim 200$ s. The deviation from the linear fit does not exceed 12% within the first 100 s, over which the profile amplitude varies by two orders of magnitude.

The result of the two parameter linear fit, $I_f = a + bt$, in the interval $0.5 < t^{1/3} < 5$ is: $b= 1.07$, $a = 1.35$ (the standard $\chi^2$ errors are meaninglessly small in this kind of fit due to correlations between bins), with $a$ corresponding to the value of the profile extrapolated to $t = 0$. The extrapolated value exceeds the actual profile value at $t = 0$ which is 1 by definition. This difference is mainly caused by the finite width of the first bin (see Fig.1b)

In order to test the $\exp(-(t/t_0)^{1/3})$ law, I varied the power index $\alpha$ in a $\exp(-(t/t_0)^\alpha)$ law, rescaled the abscissa to $t^\alpha$, and repeated the linear fit (the profile was represented by only 10 bins in this test in order to avoid a strong correlations between bins). The result of a $\chi^2$ test is $\alpha = 0.344 \pm 0.025$ at the 90% confidence level, which is sufficiently close to 1/3 that it is reasonable to suppose that $\alpha$ could be exactly 1/3.

I checked the behavior of the peak aligned profile for solar flares using an arbitrary sample of 290 flares from the same BATSE database. Fig. 2 presents the average time profile of solar flares together with the general profile of GRBs in coordinates: $-\log(-\log(I/I_0))$ vs $\log(t)$, where $I_0$ is the average intensity of the profile extrapolated to $t = 0$ (it slightly differs from $I^p$ as mentioned above). These coordinates are convenient to demonstrate dependences such as $\exp(-x^\alpha)$ in the same way as a $\log - \log$ plot demonstrates power laws $x^\alpha$. The average profile of solar flares behaves as $\exp(-(t/t_0)^{1/2})$, where $t_0 \sim 6$ sec for the first 40 seconds after peak, then it changes power index to the same value, $\alpha = $ 1/3, as for GRBs. Unfortunately, the data cover a too short time interval to make this a strong conclusion (I used only 64 ms resolution data covering 240 seconds, 1.024 sec resolution data may extend the curve somewhat). Independently of whether the profile of solar flares actually change index to 1/3 or not, peak aligned time profiles of GRBs and of solar flares have the same type of dependences.

If one looks at time profiles of individual GRBs plotted as $\log(I)$ vs t, one finds well shaped exponential tails of the pulses with the decay time constant varying from milliseconds to approximately ten seconds. However, the $\exp(-(t/t_0)^{1/3})$ behavior is hardly a sum of these exponentials. Rather it seems that the main contribution to the average time profile, decaying slower than an exponential, comes from multiple pulses within individual GRBs.

I believe that the observed shape of the time profile could be important for understanding GRBs. However even before the $\exp(-(t/t_0)^{1/3})$ law is understood, the knowledge of average GBRs time behavior presents an opportunity to measure the effect of time dilation of weak GRBs with better precision than was possible before.

## 3. Time dilation

The $\exp(-(t/t_0)^{1/3})$ shape of the time profile of GRBs is a property which complicates the measurement of the time dilation effect. When fitting the profile with a straight line in $\log(I/I^p) - t^{1/3}$ coordinates, the resulting slope is a cubic root of the time constant and all errors are multiplied by 3 when converting to normal time scale. This problem cannot be bypassed by rescaling to linear time and it also complicates the analysis of distributions of durations (the dispersion of durations is too large). This loss of accuracy is a kind of invariant and this is the probable reason why Mitrofanov et al. (1994) did not find any time dilation effect. However, knowing the actual shape of average profiles we can not only detect the time dilation but also to measure *the dependence* of the time dilation on the brightness of GRBs with satisfactory accuracy.

I measured $t_0$ for different brightness intervals using linear $\chi^2$ fits in $\log(I/I^p) - t^{1/3}$ space. The results are presented in Fig.3. It includes two sets of samples – one set with wide brightness intervals (7 samples, 70 events in each sample on average, except for the brightest sample which includes only 31 events) and a second set with narrow brightness intervals (23 samples, 20 events in each on average). The purpose of the narrow brightness range sampling was to get an independent estimate of errors. The errors estimated from the $\chi^2$ fitting procedure are not sufficiently reliable (usually being underestimated) due to correlations between neighboring time bins. The rms errors were extracted from root mean square deviation of $t_0$ iteration power law fit to 23 narrow brightness range samples. These errors were then applied to wide brightness range samples by reducing the errors assuming that the relative rms error of a sample is proportional to $N^{-1/2}$, where $N$ is the number of events in the sample.

The time dilation effect is slightly larger than found by Norris et al. (1994, 1995). Relative to the brightest sample, the time dilation i.e. the average ratio of $t_0$ for the brightest sample to $t_0$ for 3 weakest samples, is $3.2 \pm 0.7$ (The error is associated mainly with the brightest sample which includes only 31 events).

Fig. 3 indicates that the dependence of $t_0$ might be flatter in the dimmer part of the peak intensity range than in the brighter part (in the cosmological interpretation this could mean that all weak bursts are near the edge of their redshift distribution). However the statistical significance of such a flattening is small and the distribution can be fitted with a simple power law with a satisfactory $\chi^2$. I performed a power law fit to the wide brightness range samples using the rms errors obtained in the procedure described above. The result of such a fit is $\beta = -0.24 \pm 0.035$ where $\beta$ is the power index in the dependence $t_0 = (0.36 \pm 0.04)\text{s} \times (I_{1024}^p/60000 s^{-1})^{-\beta}$, $I_{1024}^p$ is the peak intensity on the 1024 ms integration time scale in units of counts per second. Note that the time dilation effect is detected in terms of a nonzero $\beta$ at a $7\sigma$ level.

These results were obtained using count rates for a wide energy band 25 – 1000 keV. However, GRBs are known to have shorter durations at high energies relative to low energies (e.g., Mitrofanov et al. 1994). I checked the dependence on energy band measuring $t_0$ for BATSE channels 1 and 2 (25 keV – 100 keV) and channels 3 and 4 (100 keV -1000keV) separately. The dependence is moderate: the $t_0$ for the softer energy band is 1.5 times larger than for the hard energy band for both bright and dim bursts.

The dependence of $t_0$ on energy band means that if we adopt the cosmological interpretation, the measured time dilation effect is slightly underestimated. Spectra of weak bursts should be redshifted (and they actually are softer, Paciesas et al. 1991, Mitrofanov et al. 1994) and a soft longer component of bursts would move below threshold. Thus what we measure for dimmer bursts is their shorter component. The possible correction for this effect is $+25 - +40$ % (taking into account the width of the energy band of channels 1 and 2). The index $\beta$ then becomes close to 0.3 and the maximal measured time dilation should be corrected to $\sim 4 \pm 0.9$. Of course, this correction makes sense only in the cosmological scenario.

## 4. Discussion

The average peak aligned profile of GRBs obeys a $\exp(-(t/t_0)^{1/3})$ law for a wide range of durations and amplitudes with good accuracy. This could be due to something more fundamental than a coincidence resulting from adding exponentials with different decay times. As I already mentioned, the main contribution to the average time profile comes from chaotic repeating pulses rather than the exponential decays of main pulses. This means that the law should be an intrinsic feature of some stochastic process responsible for the time behavior of GRBs. One possible approach to the problem is to study a number of simple stochastic toy models in the style of Rosner & Vaiana (1978).

It is also reasonable to suppose that the similarity in time behaviour between solar flares and GRBs is not a coincidence. It was pointed out in many works (e.g. Dennis 1988, Hurley 1989) that solar flares and GRBs have similar temporal and spectral properties. One can conclude from Fig.2 that GRBs just is a cleaner variant of the same class of behavior. This could mean that GRBs are associated with the same class of processes as solar flares i.e. magnetic reconnection. Magnetic reconnection can in principle be responsible for GRB's light curves not only in the case of a galactic origin of GRBs. A cosmological scenario necessarily includes a relativistic expanding fireball of a relativistic jet (e.g. Piran 1994). Both the fireball and the jet can easily be magnetically dominated, and an essential fraction of the total energy can be carried by the magnetic field. If the field is turbulent (there exist models of fireballs with turbulent field growth, see Meszaros & Rees 1994) it will reconnect during the expansion releasing energy and the light curve of a GRB will just reflect the history of these reconnections.

Concerning time dilations I believe that at the present statistical level they are consistent both with the cosmological scenario and with the existence of an anticorrelation between intensity and duration. The only way to confirm or to rule out the cosmological interpretation of time dilations is to use a larger statistical sample. The important point is that probably the necessary amount of statistics already exists in old archives.

The cosmological scenario implies a $z \sim 0$ region where there are no time dilations (unless there existed a strong evolutionary effect that completly suppressed GBRs in this region). This means that the time dilation curve should be flat for the brightest range of GRBs. The estimated time dilation is already large and probably the brightest sample in Fig. 3 should be close to $z \sim 0$ region.

The statistics which I utilized is too poor to confirm this. It covers only 2 years of observations while the total observation time of GRBs with all instruments (PVO, WATCH-GRANAT, KONUS etc.) is a few times larger. All data since the late 1960s could be utilized because the sensitivity of the instrument does not matter in this case as only light curves of strongest bursts are of interest. If a reliable signature of a $z \sim 0$ region will be detected in the dependence of $t_0$ vs peak intensity, it would leave almost no doubt that the cosmological interpretation is correct. If, on the contrary, the strongest bursts demonstrate a monotonic decrease of the $t_0$ then the time dilation between strongest and weakest bursts will exceed some reasonable cosmological limit and the effect of time dilations must require some other explanation.

I would like to thank Roland Svensson for the excellent working conditions he provided during my visit to Stockholm Observatory where this work has been performed and for valuable discussions and suggestions. I also thank Juri Poutanen, Marek Sikora, Stefan Larsson and Claes-Ingvar Bjornsson for stimulating discussions and assistance. This work is supported by the Swedish Institute and the Swedish Natural Science Research Council.

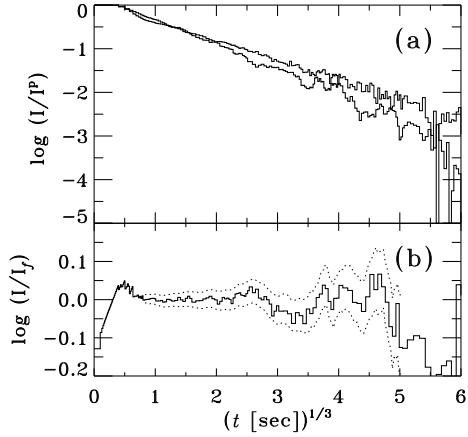

Fig. 1.— (a) Average peak aligned profiles for two brightness groups ($I^p_{1024} > 7800\text{s}^{-1}$, 95 GRBs, *thick solid curve*, and $860\text{s}^{-1} < I^p_{1024} < 3000\text{s}^{-1}$, 218 GRBs, *thin curve*). (b) Residuals of the linear fit to the average profile for 409 GRBs rescaled to $t_0 = 1$. Dotted line indicates the $1\sigma$ envelope. The straight line fragment on the left represents the first 64 ms bin assuming the count rate distribution within the bin to be uniform in $t^{1/3}$.

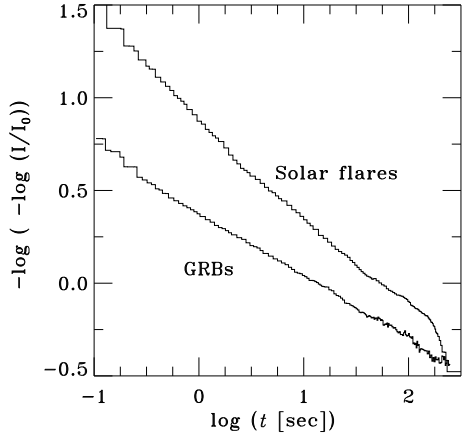

Fig. 2.— Average peak aligned profiles for 409 GRBs and for 290 solar flares in coordinates chosen to display an $\exp(-t^\alpha)$ dependence as a straight line with slope $\alpha$. $I_0$ is the profile intensity extrapolated to $t = 0$. This removes uncertainties associated with the finite width of the first bin. The cutoff at large $t$ corresponds to the 240 s duration of data records.

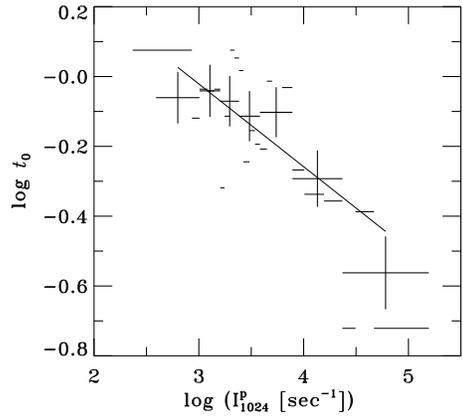

Fig. 3.— Time constant, $t_0$ versus peak intensity, $I^p_{1024}$, in log − log coordinates. Plus signs represents $t_0$ for samples of wide brightness ranges. The rms errors were estimated from the dispersion of $t_0$ for samples with narrow brightness ranges (*thin lines*).

6